\documentclass[aps,prl,twocolumn,preprintnumbers,groupedaddress,superscriptaddress,floatfix,tightenlines,reprint]{revtex4-1}
\usepackage{mathrsfs,natbib}
\usepackage{graphics,epsfig,color, graphicx}
\usepackage{verbatim}
\usepackage{bm}

\usepackage{graphicx,epsf,amssymb,bbm,amsbsy,amsfonts,amssymb,amsmath}
\usepackage{hyperref}

\newcommand{\eq}{\begin{equation}}
\newcommand{\eqe}{\end{equation}}

\def\tr{\text{tr}\,}

\newcommand{\eqa}{\begin{eqnarray}}
\newcommand{\eqae}{\end{eqnarray}}

\def\Z{{\mathbb Z}}
\def\R{{\mathbb R}}

\usepackage{verbatim}   
\usepackage[all]{xy}
\usepackage{bm}  
\def\Dslash{{\rlap{\raise 1pt \hbox{$\>/$}}D}}
\def\Pslash{{\rlap{\raise  1pt \hbox{$\>/$}}\,\partial}}

\hypersetup{
    colorlinks=true,       
    linkcolor=red,          
    citecolor=blue,        
    filecolor=magenta,      
    urlcolor=blue           
}

\begin{document}

\title{Quantum distillation in   QCD}


\author{Takuya Kanazawa}
\email{tkanazawa@nt.phys.s.u-tokyo.ac.jp}
\affiliation{Research and Development Group, Hitachi, Ltd., Kokubunji, Tokyo 185-8601, Japan}

\author{Mithat \"Unsal}
\email{unsal.mithat@gmail.com}
\affiliation{Department of Physics, North Carolina State University, Raleigh, NC 27695, USA}

\begin{abstract}
We propose a grading protocol which assigns global symmetry  associated   phases to states in the Hilbert space. 
 Without modifying  the  Hilbert space, 
 this  changes  the state sum, a process  that we call quantum distillation.   
 We describe  the image of quantum distillation 
 in terms of (non-dynamical)  flavor holonomy dependence of (dynamical) gauge holonomy potentials, in  
  QCD with $N_f=N_c$ fundamental and one massive adjoint fermion on  $\R^3 \times S^1$. The compactified  theory  possesses an  exact zero-form  color-flavor center   symmetry   for a  special  choice of  flavor holonomy (under which Polyakov loop is charged), despite the absence of one-form center-symmetry. We prove 
  that  the  CFC symmetry is stable at small-$\beta$.
  This is the   opposite of the high-temperature limit of thermal  theory  and a dramatic  manifestation of quantum distillation. 
 We show   chiral symmetry  breaking at small $S^1$ and  that the vacuum structure of the theory on $\R^4$ and   $\R^3 \times S^1$ are controlled by the same  mixed 't Hooft anomaly condition. 
\end{abstract}


\maketitle

\section{\bf Introduction }
Consider a generic QFT on a $d$-dimensional space-manifold ${\cal M}$. 
The dimension of the Hilbert space of such a theory  scales exponentially with the volume, $e^{ {\rm Vol}({\cal M})}$. 
It is a monstrous structure, and  for different purposes, one cares very little with the detailed knowledge of the full space. 
For low energy physics,  one cares solely about ground states and low lying states.  In the theory of phase transitions,  
the interest is in the growth of the  density of states $\rho(E)$   
and high energy states become crucial. 

Consider a thermal partition function associated with an asymptotically free  or super-renormalizable QFT, 
${\cal Z}(\beta)= \tr [e^{- \beta H}]$, where $\beta$ is inverse temperature.  In the $\beta \rightarrow \infty$ limit, this receives  dominant contributions from  ground states and low lying states. However, in this limit, these QFTs are often strongly coupled and not amenable to analytic treatment.  
As $\beta \rightarrow 0 $,  they may become weakly coupled, but the state sum receives contributions from every state on the same footing. It is an utterly contaminated quantity. In this regime, it is  impossible to isolate and understand the role of ground states and low lying states.  Relatedly, the theory often moves to a different phase at some $\beta < \beta_c$ due to  phase transitions associated with  singularities of partition function ${\cal Z}(\beta)$.

On the other hand, despite the fact that the full Hilbert spaces of supersymmetric theories are also equally complicated,  
one can often succeed in constructing  graded state sums, e.g.,  supersymmetric  Witten index  \cite{Witten:1982df},  which receives  contributions only from  ground states,  ${\cal Z}(\beta)= \tr [e^{- \beta H} (-1)^F]$.   This is a  Bose-Fermi paired graded state sum, in which  positive energy states $E>0$ cancel pairwise thanks to supersymmetry (assuming that the spectrum is discrete). 

In this work, we are after something  ambitious. We would like to construct a graded state sum in certain four-dimensional  QCD-like theories, such that the cancellation among states can prevent all phase transitions as one moves from large-$\beta$ to small-$\beta$.  We would like to achieve this without changing the theory, its Hilbert space, and without adding   center-stabilizing double-trace deformations \cite{Shifman:2008ja, Unsal:2008ch}.

The goal is  to construct a graded  state  sum which can avoid all singularities in ${\cal Z}(\beta)$. One may think that this is  impossible in a non-trivial non-supersymmetric theory due to lack of supersymmetry. 
 However, we will  prove  otherwise in  interesting theories,  including $N_f=N_c$ QCD  with an  extra heavy degree of freedom.  
  We refer to the protocol of changing the state sum, which  reduces the sum over the Hilbert space  ${\cal H}$ down to a much smaller subset of states,  ${\rm Distill}[{\cal H}]$,  as {\it quantum distillation.}   The origin of this idea in two-dimensional QFTs  can be traced to  \cite{Dunne:2012ae,  Cherman:2014ofa, Sulejmanpasic:2016llc, Dunne:2018hog}. The spectral miracles in  adjoint QCD in 4 dimensions  \cite{ Basar:2013sza, Basar:2014jua, Cherman:2018mya} can also be interpreted in this language.

\section{\bf Quantum distillation in QCD(F/adj)} 
Consider  QCD with $N_f=N_c $ flavors of fundamental massless Dirac fermions 
$\psi^{a}$ 
and one   adjoint massless Weyl  
fermion $\lambda$  
 with Euclidean Lagrangian:
\begin{align} 
{\cal L}= \frac{1}{2g^2} \tr F_{\mu \nu}^2 + \sum_{a=1}^{N_f}  \overline{\psi}_a \gamma_\mu D_{\mu} \psi^a +   2 \;\tr  \overline{\lambda} \bar \sigma_\mu D_{\mu}  \lambda
\label{Lag}
\end{align} 
The global  symmetry of the theory  $\bf G_{non-ab}$   that acts faithfully on the Hilbert space is:
\begin{equation}
{SU(N_f)_L \times SU(N_f)_R \times U(1)_V  \times U(1)_{A_D} \times \mathbb{Z}_{2\mathrm{gcd}(N_c,N_f)}\over \mathbb{Z}_{N_c}\times (\mathbb{Z}_{N_f})_L\times (\mathbb{Z}_{N_f})_R\times (\mathbb{Z}_2)_\psi}.
\label{symmetry}
\end{equation}
where $U(1)_{A_D}$ is the diagonal subgroup of the two classical chiral  $U(1)_A$ acting on $\psi^{a}$  and  $\lambda$, respectively.  
$\Z_{2 {\rm gcd}(N_c,N_f)}  $ is the discrete remnant of classical chiral symmetry. 
By turning on a mass $m_\lambda > 0$  for adjoint fermion, the theory can be reduced to QCD at low energy. There are two  reasonable scenarios for the chiral symmetry breaking. {\it  1)} The whole axial part is broken as in QCD.  
 {\it  2)}  All  but $U(1)_{A_D}$ axial symmetry  is broken as in $N_f=N_c$ SQCD \cite{Affleck:1983mk, Aharony:1995zh}.
 Of course, once a mass term  $m_\lambda $ is turned on (which is our  interest in this work),  $U(1)_{A_D}$   is lost  and  there is no difference between these two scenarios.

We would like to study the dynamics of the theory by using a generalized partition function: 
\begin{align} 
{\cal Z}(\beta, \epsilon_a)  = \tr \Big[ e^{-\beta H} (-1)^F  \prod_{a=1}^{N_f} e^{i \epsilon_a Q_a}  \Big] 
\label{GPF}
\end{align} 
where $Q_a$ are the charges corresponding to  Cartan generators of $U(N_f)_V$  and $(-1)^F$ is grading by fermion number. These operators commute with the Hamiltonian $H$, and do not alter the Hilbert space ${\cal H}$. In path integral formulation, this corresponds to the   boundary conditions 
 \begin{align}
 \lambda(\beta) &= + \lambda(0), \qquad \psi(\beta)= +    \psi(0)  \overline \Omega_{F} e^{i \pi} \cr 
 \Omega_{F}&= {\rm diag} \big( e^{i \epsilon_1}, \ldots, e^{i \epsilon^{}_{N_f}} \big)
 \label{bc}
\end{align}
where  $\Omega_{F}$ is a flavor twist, and  $ e^{i \pi} $ is factored  out for convenience.  $\Omega_{F}$  may  also be viewed  as an imaginary chemical potential,  and Hilbert space grading  and sum provide a physical interpretation for it.

 In the presence of fundamental fermions,  it is well-known that the one-form (and zero-form in compactified theory)  center-symmetry  is explicitly broken, and Wilson loops (Polyakov loops for zero-form)   are no longer good order parameters, see  \cite{Gross:1980br, Kapusta:2006pm,Laine:2016hma}.     
  However,      consider a  $U(N_f)_V$ flavor twisted boundary condition on fermions, setting $N_f=N_c$:
\begin{align}
 \Omega_F^0 = \textrm{diag}
(1, \omega, \cdots, \omega^{N_f-1}), \qquad \omega = e^{2\pi i /N_f} . 
\label{flavor-hol-0}
\end{align}
where  $\psi (x_4+ \beta) = -  \psi (x_4)  \overline \Omega_F^0$. Under a gauge transformation aperiodic up to an element of the center, 
$\psi^G (x_4+ \beta) = -  \psi^G (x_4) \overline \Omega_F^0  $, the 
 boundary conditions  maps  into $\psi (x_4+ \beta) = -  \psi (x_4) \overline \Omega_F^0    \omega $. Since this boundary condition is  different from the original one,  this is  a non-invariance of partition function and explicitly breaks zero-form center-symmetry.  On the other hand, 
  $\omega \overline \Omega_F^0 $ is a cyclic permutation of $
 \overline \Omega_F^0$ and can be brought to the original boundary conditions by using cyclic permutation matrix   $(S)_{a, b} = \delta_{a+1, b}  \in \Gamma_S \subset  SU(N_f)_V$, and turn the combined operation into a  genuine global symmetry.

   For general $N_f, N_c$,  the  compactified theory  possesses   an {\bf exact}    
  $\Z_{{\rm gcd}(N_f, N_c)}$  zero form color-flavor  center  (CFC) symmetry   under which Polyakov loops are charged,     despite the fact that it does not possess  a one-form center symmetry under which Wilson loops are charged   \cite{Cherman:2017tey, Iritani:2015ara, Poppitz:2013zqa}. We will see manifestation of this fact explicitly in holonomy potential. 
  Therefore, we can examine the phase structure of these theories according to the CFC symmetry and pose questions about analyticity of graded partition function   as a function of $\beta$. The CFC  is one of the major players in our construction. 
  
The price one pays for keeping an exact 0-form center-symmetry is the reduction of the non-abelian chiral symmetry to a subgroup which commutes with $\Omega_F^0$, the  maximal abelian sub-group ${\bf G}_{\rm max-ab}$
 \begin{align}
{U(1)^{N_f-1}_L  \times U(1)^{N_f-1}_R \times U(1)_V \times U(1)_{A_D}   \times \mathbb{Z}_{2\mathrm{gcd}(N_c,N_f)}\over \mathbb{Z}_{N_c}\times (\mathbb{Z}_{N_f})_L\times (\mathbb{Z}_{N_f})_R\times (\mathbb{Z}_2)_\psi}.
\label{MAG}
\end{align}
This is the exact symmetry at any finite-$\beta$. When $\beta$ is much larger than the strong length scale, 
\eqref{symmetry} should be viewed as approximate symmetry and ultimately recovered at $\beta \rightarrow \infty$ limit.

\section{\bf  Gauge holonomy potentials without and with flavor holonomy twists}
In the  thermal case, as $\beta \rightarrow 0$, the theory  moves to a chirally restored phase. 
This regime is sometimes called deconfined, but strictly speaking, center-symmetry  is not present due to  the fundamental matter fields  with thermal boundary conditions. 
 In the small-$\beta$ regime, the potential for gauge holonomy $\Omega= e^{i \oint a_4 dx_4}$   is well-known   \cite{Gross:1980br,Weiss:1981ev}:
\begin{eqnarray} 
V_\text{1-loop, th.}(\Omega)   
& =&  \frac{2}{\pi^2 \beta^4} \bigg\{ \sum_{n=1}^{\infty} \frac{1}{n^4}    [-1 + (-1)^n ]    |\tr (\Omega^n)|^2    \cr \cr
& & + N_f  \sum_{n=1}^{\infty} \frac{ (-1)^n }{n^4}       \left[    \tr (\Omega^n)+ {\rm c.c.}  \right]  \bigg\}
\label{1-loop-thermal}
\end{eqnarray} 
where the contributions are respectively from $a_\mu, \lambda, \psi_a$.   

The leading order free energy, which is found by minimizing this potential, is given by (see e.g., \cite{Kapusta:2006pm, Laine:2016hma, Philipsen:2012nu})
\begin{align}
{\cal F}_{\rm th} (\beta)&= -  \frac{\pi^2}{90}   \frac{V_3}{\beta^4}  \Big[  \underbrace{2(N_c^2-1)}_{ \rm gluons} +  \frac{7}{8}     ( \underbrace{  2 (N_c^2-1)}_{\rm adj.  Weyl.}   +   \underbrace{ 4 N_f  N_c }_{\rm fund. D.}) \Big] 
\end{align} 
which can be interpreted in two different ways following quark-hadron duality \cite{Shifman:2000jv, Shifman:2005zn}. The standard (microscopic) interpretation is  the  usual  Stefan-Boltzmann  free energy of asymptotically free quarks and gluons.
 The   macroscopic interpretation is in terms of the hadronic density of states  in the Hilbert space ${\cal H}$. The inverse Laplace transform of the partition function (for $N_f\sim N_c$) ${\cal Z}(\beta) \sim e^{- \beta {\cal F}_{\rm th}}  \sim e^{a N_c^2   V_3 /\beta^3 }  $  ($a$ is a pure number)  is just the density of states of hadrons in the spectrum, 
$\rho_{\rm SB}(E) \sim  e^{E^{3/4} N_c^{1/2} {(a V_3)}^{1/4}} $, where SB stands for the Stefan-Boltzmann growth.  
The idea of quantum distillation is to create sufficient destructive interference in the state sum such that an effective  density of states (the one associated with ${\rm Distill}[{\cal H}]$)  is not strong enough to change the phase of the theory as $\beta$ is reduced.

The gap between the minimum and maximum of the holonomy potential \eqref{1-loop-thermal}  is $ \Delta V \sim O(N_c^2)$. 
 The configuration $\Omega= {\mathbbm{1}}_{N_c}$ is the minimum and governs the properties of the  thermal equilibrium state.  In particular, the center-symmetric configurations    are close to maxima.   All three terms in the  GPY potential \eqref{1-loop-thermal}  need to be defeated in order  for the quantum distillation idea to work!

The effect of grading in state sum  \eqref{GPF} maps into non-dynamical flavor holonomy dependence in the gauge holonomy potential. 
At one-loop order, we find: 
 \begin{align}
 V_{\textrm{1-loop}, \Omega_F} & =  V^{\rm  gauge}_{\textrm{1-loop}} + V^{\rm \lambda}_{\textrm{1-loop}}  +  V^{\rm \psi}_{\textrm{1-loop}, \Omega_F}\,,\notag\\
 V^{\rm  gauge}_{\textrm{1-loop}} + V^{\rm \lambda}_{\textrm{1-loop}}  & = (-1 +1) 
 \frac{2}{\pi^2 \beta^4}  \sum_{n=1}^{\infty}   \frac{1}{n^4}   
        |\tr (\Omega^n)|^2  =0 \,, \notag\\
V^{\rm \psi}_{\textrm{1-loop}, \Omega_F} & =
 \frac{2}{\pi^2 \beta^4}  \sum_{n=1}^{\infty}  \frac{(-1)^n}{n^4}    
   \left[ \tr (\overline{\Omega}_F^n)  \tr (\Omega^n) + \text{c.c.}\right]  
  \label{one-loop-A}
\end{align}
The effect of gauge fluctuations is undone by one massless adjoint fermion (but recall that we will   
make adjoint fermion heavy and reduce the low energy  theory to QCD.)
 This is true to all orders in perturbation theory,  because the  subclass of  Feynman diagrams (in loop expansion) composed {\it solely} of $a_{\mu}$ and $\lambda$  is identical to   ${\cal N}=1$ SYM, hence a potential cannot be generated from this class.   Furthermore,  at one and two-loop order, $\lambda$  and $\psi^a$ are decoupled in loop expansion.  
 Starting at three-loop order, there are diagrams involving both fundamental and adjoint fermions. However,   the  CFC  realization  can be determined reliably  with two-loop knowledge.

 For $\Omega_F= \Omega_F^0$ given in   \eqref{flavor-hol-0},  since $\tr (\Omega_F^0)^n = 0$ unless $n=N_fk, \; k\in\mathbb{Z}^{+}$, 
  a number of remarkable effects take place. {\bf a)}  We observe the manifestation of the exact CFC symmetry at one-loop order.  
The potential becomes  invariant under  $ {  \Z_{{\rm gcd}(N_f, N_c)}} $  symmetry  since all terms of the form 
  $  \tr (\Omega^n), n \neq N_f k $ vanish. 
{\bf b)}  The gap  between the minimum and maximum of the potential, for $N_f\sim N_c$,   becomes $ \Delta  V_{\textrm{1-loop}, \Omega_F^0}  \sim O(1/N_c^2)$, i.e., the one-loop potential is  extremely frustrated  and it   collapses!  At one-loop order, one obtains    an   exponentially increasing number of degenerate minima,   
   ${\frak {N}}_{\rm min} (N_c) \approx   \frac{2^{2N_c-1}N_c^{-3/2}}{\sqrt{\pi}}$.  In the $N_c \rightarrow \infty$ limit, the gauge holonomy potential vanishes at one-loop order and the set of minima becomes continuous similar to supersymmetric theories. {\bf c)} 
  The graded free energy at the level of one-loop analysis  (for $N_f \sim N_c$)  takes the form:
  \begin{eqnarray}
{\cal F}_{\Omega_F^0}(\beta)&=& -  \frac{\pi^2}{90}   \frac{V_3}{\beta^4}  \Big[  \underbrace{2(N_c^2-1)}_{ \rm gluons} -   \underbrace{  2 (N_c^2-1)}_{\rm adj.  Weyl.}   +   \underbrace{ \frac{7}{2} \frac{1}{N_c^2}  }_{\rm fund. D.}) \Big]  \cr
&=&  -  \frac{\pi^2}{90}   \frac{V_3}{\beta^4}    \left[   \frac{7}{2} \frac{1}{N_c^2}   \right]   \longrightarrow{   0, } {\qquad N_c \rightarrow \infty\,. }  
\end{eqnarray}  
This is the effect of quantum distillation.  It   is as if there is merely  $\frac{1}{N_c^2}$ quark degree of freedom in the system instead of $\sim N_c^2$ bosonic and $\sim N_c^2$  fermionic degrees of freedom!   
 The corresponding scaling of the graded partition function  is  
$ {\cal Z}(\beta) \sim e^{\frac {a}{N_c^2}   V_3/\beta^3   } $ and 
and the effective density of states  of the hadronic states  in   ${\rm Distill}[{\cal H}]$  is given by 
$\rho_{\Omega_F^0} (E) \sim  e^{ \frac{1}{N_c^{1/2}} E^{3/4}  { (a V_3)}^{1/4}}$.

In order to determine CFC-realization, we need a two-loop result in the presence of the grading.  
To this end,   we  generalize    earlier  thermal studies  of Refs.~\cite{KorthalsAltes:1993ca,KorthalsAltes:1999cp} and \cite{Guo:2018scp} to incorporate flavor-holonomy.  We use Eq.(17) in  
\cite{KorthalsAltes:1999cp} and    Eq.(5.11) in \cite{Guo:2018scp}.  
These two results at at fist sight look  different, but due to  non-trivial Bernoulli polynomial identities, they are actually the same.  

  The two-loop potential in the presence of the  flavor-twisted boundary conditions is \eqref{bc}: 
\begin{eqnarray}
	V_{ \text{2-l}, \Omega_F}^\psi & =& \frac{g^2}{\beta^4}
	\frac{3}{\pi^4}\left\{-\frac{N^2_c-1}{8N_c}\sum_{n=1}^{\infty}\frac{(-1)^n}{n^4}[\text{Tr}(\overline{\Omega}_F^n)\text{Tr}(\Omega^n)+\text{c.c.}]  \right.
	\cr 
&&	 \left.  + \frac{N_f}{24}\sum_{n=1}^{\infty}\frac{\big| \text{Tr} (\Omega^n) \big|^2}{n^4} \right\} 
	\label{V2psi}
\end{eqnarray} 
Again, in the  $\Omega_F= \Omega_F^0$ background,  in the first sum,  all but $n=N_f k$ terms vanish, and the sum is $O((g^2N_c)/N_c^2)$ without altering any of the results of one-loop analysis.  In the large-$N_c$ Veneziano limit \cite{Veneziano:1976wm}  of QCD(F/adj),  the combined one- and two-loop potential takes a simple and beautiful   form
\begin{align} 
 V_{\textrm{1-l}, \Omega_F^0}  + V_{\textrm{2-l}, \Omega_F^0} =
    +   \frac{g^2N_f}{ 8 \pi^4 \beta^4} 	\sum_{n=1}^{\infty}\frac{\big| \text{Tr} (\Omega^n) \big|^2}{n^4}  \,, 
	\label{V12largeN-1}
\end{align} 
This is one of the main results of this work. Fundamental fermions with $\Omega_F^0$-twisted boundary conditions at two-loop order  is capable of stabilizing the  CFC-symmetry, by inducing center stabilizing   double-trace terms at two-loop order. In this sense, they behave  similar to  adjoint fermions with periodic boundary conditions \cite{Kovtun:2007py}. 
  The minimum of the gauge-holonomy potential is  now at 
\begin{align}
\Omega \big|_{\rm min}&=   \omega^{- {(N_c-1)/2} }\text{diag}(1,\omega,\cdots,\omega^{N_c-1}), 	\label{centersym0}%
\end{align}
  The stability  of center symmetry is robust to  all loop orders in perturbation theory and non-perturbatively in the weak coupling small-$\beta$ regime.   The stability  is the image of the  quantum distillation over the Hilbert space in the gauge holonomy potential. 

Turning on  $m_\lambda$, the  balance between the gauge fluctuation and adjoint fermion breaks  in favor of CFC-breaking.    However, since the two-loop effects of the fundamental fermions is also CFC-stabilizing,  the  full $\Z_{N_c}$ center symmetry can be kept intact provided   $m_\lambda \leq   m_\lambda^*= \frac{(g^2N_c)^{1/2} }{  N_c \beta  \pi}$.   
Therefore, 
provided  $\Lambda  \ll    m_\lambda \leq   m_\lambda^* $ (which is easily  achieved at finite $N_c$), the IR theory is essentially QCD(F) both on $\R^4$ as well as on $\R^3 \times S^1$.  
In other words, in this window the adjoint fermion does not decouple from the holonomy potential, but it does decouple  from other aspects of the long distance physics, such as chiral Lagrangian.

Therefore, the net effect of quantum distillation is the $1/N_c^4$ suppression in free energy  (at least at two-loop order)  and the suppression of the  hadronic density of states 
 \begin{align}
\rho_{\rm SB}(E) &\sim  e^{E^{3/4} N_c^{1/2} {(a(g^2N_c) V_3)}^{1/4}}  \cr
&\mapsto \rho_{\Omega_F^0} (E) \sim  e^{ \frac{1}{N_c^{1/2}} E^{3/4}  {(a (g^2N_c) V_3)}^{1/4}}
\end{align}
where $a (g^2N_c)$ is a series expansion in 't Hooft coupling. 
It is highly plausible, similar to \cite{Cherman:2018mya},  that the  growth ${\rm exp} (V_3^{1/4}E^{3/4}) $ expected in a standard local 4d theory in spatial volume $V_3$ cancels and   $\rho_{\Omega_F^0} (E)$ turns out to have a scaling associated with a 
two-dimensional QFT in the  $N_c \rightarrow \infty$ limit.

In the small circle regime where CFC remains  unbroken,   the chiral symmetry breaks spontaneously  by the condensation of monopole flux operators as described in \cite{Cherman:2016hcd}.   In appendix,  we present   this mechanism  in 
operator  formalism in contrast with  \cite{Cherman:2016hcd}. 
 Therefore, we conjecture that the small-$S^1$ chirally broken phase of QCD(F/adj) with $\Lambda \ll m_\lambda <  m_\lambda^*$ is adiabatically connected to the strong coupling regime, and quantum distillation achieves its goal, the continuity of the partition function. 
We cannot prove this highly plausible conjecture, but instead we can prove   weaker  statement that 
 the vacuum structure of the theory on any size   $\R^3 \times S^1$ and $\R^4$ is  controlled by the same    mixed 't Hooft anomaly condition.

\section{\bf Mixed anomaly and persistent order}  
On $\R^4$, the 
 faithful  symmetry 
  of the $\Lambda \ll m_\lambda <  m_\lambda^*$   theory includes   $G_1= SU(N_f)_V/ \Z_{{\rm gcd}(N_f, N_c)}$ because   $ \Z_{{\rm gcd}(N_f, N_c)} \subset\Z_{ N_c}  $  
 is part of gauge structure, and therefore, is not a symmetry.  The theory also has a  $G_2=\Z_{2N_f}$ chiral symmetry, 
 which  resides in continuous chiral symmetry.  By using the techniques of \cite{ tHooft:1979rat, Gaiotto:2017yup, Gaiotto:2014kfa, Tanizaki:2017qhf, Shimizu:2017asf, Komargodski:2017smk}, it is possible to show that there is 
  a mixed anomaly between  these two symmetries.

  Introducing the  $SU(N_f)$ gauge field $A$,  the theory becomes $SU(N_f) \times  SU(N_c)$  quiver theory (bifundamental QCD), which has a  genuine  $ \Z_{{\rm gcd}(N_f, N_c)}$ one-form symmetry.  To gauge    
  the correct symmetry, $G_1$,  the one-form symmetry must  also  be gauged, and  for this purpose, we introduce a two-form gauge field $B$. 
 In quiver theory, only  $ \Z_{2   {\rm gcd}(N_f, N_c) } \subset     \Z_{2N_f}$ chiral symmetry is  present. 
Generalizing   Ref.~\cite{Tanizaki:2017qhf} to arbitrary $N_f$ and $N_c$,  we observe that under the  discrete $\chi$S transformation $h$,  the partition function fails to be invariant 
 \begin{align}
{\cal Z}(h(A, B))  = e^{        - i  { 2 {\rm lcm}(N_f, N_c)   \over 4\pi } \int 
  B  \wedge B   } {\cal Z}(A, B)   
  \label{poly}
 \end{align}
 provided 
  \begin{align}
 \frac{  2 {\rm lcm}(N_f, N_c)}  { \left( {\rm gcd} (N_f, N_c) \right)^2 }\in   {\mathbb Q \backslash \mathbb Z}
 \label{anomalycon}
  \end{align}
  For all $N_f=N_c \geq 3$,    there is a mixed anomaly  since  $ \frac{2}{N_c} \in   {\mathbb Q \backslash \mathbb Z}$ in agreement with 
 \cite{Tanizaki:2017qhf}.

This mixed 't Hooft anomaly implies that a unique, gapped   ground state  is  impossible.  Therefore,  either {\it a)}  $SU(N_f)_V/ \Z_{{\rm gcd}(N_f, N_c)}$ or  {\it b)}     $\Z_{2N_f}$  or {\it c)}  both  are spontaneously broken, 
or 
{\it d)} IR theory has to be a CFT. 
In the present case, we can rule out  {\it a)}   and  {\it c)}    because of the Vafa-Witten theorem, which asserts that in vector-like theories, vector-like global symmetries cannot be spontaneously broken  as long as one assures  positivity of the measure \cite{Vafa:1983tf}. Fortunately, the  grading \eqref{GPF} respects this positivity. 
For QCD-like theories, both $\chi$SB as well as CFT options are equally reasonable. 
 From now on, we work with the assumption that   $\Z_{2N_f}$ discrete symmetry  is broken. 
As observed in \cite{Cherman:2017dwt}, there exists no order parameter which is charged under the discrete $\chi$S but not under continuous $\chi$S, (the opposite statement is not true),   implying spontaneous breaking of $SU(N_f)_A$.  Therefore, massless pions must exist in the spectrum. 

Ref.\cite{Gaiotto:2017yup} recently showed that 
a mixed anomaly between a  1-form symmetry  and 0-form symmetry persists upon compactification  on  $\R^3 \times S^1$, see  also  \cite{ Komargodski:2017smk, Shimizu:2017asf}.   This is in sharp   contrast with  mixed anomalies involving only 0-form symmetries  originally discussed by 't Hooft \cite{tHooft:1979rat} which do not survive compactification.    However, there is an  important exception for the latter \cite{Tanizaki:2017qhf}.  Assume we have  0-form symmetries $G_1 \times G_2$, where $G_1= \widetilde G_1/\Gamma$ and  gauging $ \widetilde G_1 $ turns $\Gamma$ into a genuine one-form symmetry. This is precisely the case in QCD(F/adj).   Then, a triple-mixed anomaly  persists upon compactification provided some conditions are satisfied. 

The  anomaly \eqref{poly} persists compactification on $\R^3 \times S^1$  if and only if  one uses 
a flavor twisted boundary conditions by $\Omega_F^0$ \cite{Tanizaki:2017mtm}. 
  Indeed, these boundary conditions are used in 
  \cite{Dunne:2012ae, Cherman:2014ofa,  Sulejmanpasic:2016llc, Dunne:2018hog} to preserve adiabatic continuity. 
In QCD(F/adj), the combination of center-transformation and   $\Gamma_S $  cyclic permutation subgroup of $SU(N_f)_V$,   remains as a true zero-form  symmetry of the compactified theory.  This is identified as CFC  symmetry in \cite{Cherman:2017tey}. 
 Introducing the background gauge field  for $U(1)^{N_f-1}$,  $\Z_{{\rm gcd}(N_f, N_c)} $ emerges as a 1-form symmetry. We also introduce a 2-form field $B^{(2)}$ and 1-form field  $B^{(1)}$ associated with 1-form and 0-form part of center-symmetry 
$ \Z_{{\rm gcd}(N_f, N_c)}$. Then, the  partition function fails to be invariant under  $h\in \Z_{2N_f}$ transformation as: 
\begin{multline} 
{\cal Z}_{\Omega_F^0 } ( h( A_K,   B^{(2)}, B^{(1)})) = \\
  \quad \quad  e^{  - i  { 2 {\rm lcm}(N_f, N_c)   \over 2 \pi } \int 
  B^{(2)}  \wedge B^{(1)}    } {\cal Z}_{\Omega_F^0 } ( A_K,   B^{(2)}, B^{(1)}) 
\end{multline}
Therefore, there is a triple  mixed anomaly between  $\Gamma_S$ shift symmetry, abelianized flavor symmetry $U(1)^{N_f-1}/\Z_{{\rm gcd}(N_f, N_c)}$,  and the discrete $\chi$S   $\Z_{2N_f}  \subset  U(1)_A^{N_f-1}$  provided \eqref{anomalycon} holds, the same condition as in $\R^4$.   Indeed, 
this  anomaly polynomial naturally  descends from the  one on  $\R^4$  \eqref{poly} with the substitution   $B=   B^{(2)} +   B^{(1)}  \wedge \beta^{-1} dx^4$. 
Again, the   mixed anomaly implies that the ground state cannot be  unique and  gapped.   The options are 
$U(1)_A^{N_f-1} $  chiral symmetry  breaking, CFC breaking,  spontaneous breaking of both, or a CFT behaviour at low energy. Semi-classics  (described in Appendix) proves that the first option takes place with the use of ${\Omega_F^0 }$ twisting at small-$\beta$.

\section{\bf Conclusion} Despite the fact that  we did not prove adiabatic continuity between the weak and strong coupling regimes, 
we reduced the possibilities to just a few thanks to
persistent order due to the mixed anomaly.

 On small $\R^3 \times S^1$, we showed that one can keep the 
 CFC symmetry intact with a unique  choice of boundary conditions in QCD(F/adj). One can sufficiently decouple adjoint fermions to reduce the theory to 
 real QCD(F).  The preservation of the CFC symmetry is a    manifestation of the distillation of Hilbert space,  similar to what 
 $\tr [e^{- \beta H} (-1)^F]$ achieves in supersymmetric theories.

 We showed that the twisted  boundary conditions which satisfy maximal quantum distillation is the one for which the anomaly polynomial on $\R^4$   naturally descends to   $\R^3 \times S^1$.  On small $S^1$, we proved chiral symmetry breaking rigorously with semi-classical methods. On large $S^1$, and $\R^4$, the ground states is  controlled by the same mixed anomaly structure as small $S^1$.

  Our work has interesting connections with  spectral conspiracies in non-susy theories \cite{Cherman:2018mya, Kutasov:1990sv, Dienes:1995pm},  and 
  volume independence \cite{EguchiKawaiOriginal,  GonzalezArroyo:1982hz, 
   Basar:2013sza, GonzalezArroyo:2010ss,  Perez:2014sqa} whose working examples must have an interpretation 
  in terms of quantum distillation.

Our construction creates a sign problem in the Hamiltonian  formalism, in the state sum, but not in the Euclidean path integral formulation. In this sense, it is a good sign ``problem", opposite to the notorious QCD sign problem at finite chemical potential. Therefore,  it is possible to test our analytic construction via numerical lattice simulations involving light fermions.

%
%
%
{\bf Acknowledgments.}  We are grateful to Aleksey Cherman, David Gross, Yuya Tanizaki   for discussions.  
M.\"U. acknowledges support from U.S. Department of Energy, Office of Science, Office of Nuclear Physics under Award Number DE-FG02-03ER41260.

\bibliographystyle{JHEP}
\bibliography{small_circle} 

\newpage
\appendix 

\begin{center} 
{\bf Supplemental Material} 
\end{center} 

{\bf Chiral symmetry breaking:} 
 In this short appendix, we present an operator description  of chiral symmetry breaking on 
small $\R^3 \times S^1$ in the regime where color-flavor center symmetry is unbroken and 
$\Lambda \ll m_\lambda <  m_\lambda^*$.   
We  set $N_f =N_c$ for simplicity. Our result is same as  \cite{Cherman:2016hcd},  translated to operator language, and has the virtue of providing interesting insights.

At the gauge holonomy  configuration \eqref{centersym0}, the  theory dynamically abelianize down to $U(1)^{N_c-1}$ at distances larger than inverse $W$-boson mass.  The $U(1)^{N_c-1}$ photons can be dualized to gapless scalars, $F \sim * d \sigma$, which remain gapless to all orders in perturbation theory. 
The gaplessness is protected by the  topological shift symmetry: 
\begin{align}
[U(1)_J]^{N_c -1}:  \sigma  \rightarrow  \sigma   +   \varepsilon,  \qquad  {\cal J}_{\mu} = \partial_{\mu}  \sigma  
\label{shift-mix}
\end{align}

Non-perturbatively, there are monopole-instanton effects, which come in $N_c$-types, associated with the affine root system of $\frak{ su}(N_c)$ Lie algebra   \cite{ Lee:1997vp, Kraan:1998sn}.  
 The form of the monopole operators depend on the interplay between the flavor holonomy and gauge holonomy. Without loss of generality, and with a particular choice of flavor holonomy,   the monopole operators  can be written as 
\begin{align}
 {\cal M}_i \sim  e^{-S_i}    \;  e^{ -  \frac{4 \pi}{g^2}  \alpha_i \cdot \phi +   i \alpha_i \cdot \sigma } 
      (\psi_{L}^i \psi_{Ri}),  
      \label{mon}
\end{align}

Consider a collection of $n_i$ monopoles of type-$i$ $i= 1, \ldots, N_c$ sprinkled  in between two asymptotic time slice.  Then,  the magnetic charge non-conservation is 
\begin{align}
\Delta {\bf Q}_m &\equiv {\bf Q}_m (t= \infty) - {\bf Q}_m (t= - \infty)  \cr
&= \int d^2 x  F_{12}  \Big |_{t=-\infty}^{t=+\infty}  
 =  \int_{S^2_{\infty}}  F_{12}  \cr
&= \frac{4 \pi }{g} \sum_{i=1}^{N_c} n_i \alpha_i 
\end{align} 
These charges violate {\it emergent} $ [U(1)_J]^{N_c -1}$ explicitly. 
Based on this violation, which is same as in the  old Polyakov model \cite{Polyakov:1976fu}, one may be tempted to think that  the  $\sigma$ fluctuations will be  non-perturbatively gapped due to proliferation of monopoles. 
  But the story is actually opposite!    In this background, 
 the 
  axial charge associated with ${\bf G}_{\rm max-ab}$ \eqref{MAG}  is also  not conserved:
\begin{align}
\Delta {\bf Q}^5  &= 
\sum_{i=1}^{N_f} n_i  \alpha_i 
\end{align} 
 Naively,  this would mean that ${\bf G}_{\rm max-ab}$ is anomalous. But this  is impossible since  
it lives in $\bf G_{non-ab}$, which is  manifestly anomaly free!  The resolution of this puzzle is  generalization of a mechanism discovered  by  Affleck, Harvey and Witten \cite{Affleck:1982as} in a gauge theory on   $\R^3$.  
A linear combination  of these two-charges  is  non-perturbatively  conserved: 
\begin{align} 
\Delta {\bf \widetilde Q}  \equiv \Delta \left( \frac{g}{4\pi} {\bf  Q}_m  -   {\bf Q}^5 \right)= 0
\label{conserved}
\end{align}
 Since the $\bf G_{ab}$ is the  true  microscopic symmetry,   and $[U(1)_J]^{N_c -1}$ is emergent symmetry to  all orders in  perturbation theory,   this  mechanism is present  so that the chiral charge of the fermion bilinear is  transferred  into gauge fluctuations!  In the IR theory, gauge fluctuations (dual photon field)  becomes  chirally charged. 

The  pure flux part of the monopole operators  can be combined into a diagonal component of the  matrix field, 
which can be interpreted as the {\it chiral field } of the chiral Lagrangian $ \Sigma(x) = {\rm Diag}\left( 
  e^{i \alpha_1\cdot  \sigma}, \ldots, e^{i \alpha_{N_c}\cdot  \sigma} \right)$.  $ \Sigma$ manifold is the maximal torus  of chiral symmetry, and \eqref{conserved} forbids formation of a potential on it. 
 Choosing a point on the  $\Sigma$ field manifold correspond to spontaneous chiral symmetry breaking, and IR theory is described by $S= \int_{\R^3 \times S^1}  \frac{f_\pi^2}{4}  \tr |\partial_{\mu} \Sigma|^2$. 
 Turning on a small mass for quarks breaks chiral symmetry softly,   and induce a potential $\sim m_\psi  e^{-S_0}  \tr ( \Sigma + \Sigma^{\dagger} )$.  The theory acquires a non-perturbatively induced mass gap on small $\R^3 \times S^1$. 
For details of this chiral symmetry breaking mechanism in path integral description, see  \cite{Cherman:2016hcd}.

\end{document}